\documentclass[conference]{IEEEtran}
\IEEEoverridecommandlockouts

\usepackage{cite}
\usepackage{amsmath,amssymb,amsfonts}
\usepackage{algorithmic}
\usepackage{graphicx}

\usepackage{bm}
\usepackage{multirow}
\usepackage{subfigure}
\usepackage{textcomp}
\usepackage{xcolor}
\def\BibTeX{{\rm B\kern-.05em{\sc i\kern-.025em b}\kern-.08em
    T\kern-.1667em\lower.7ex\hbox{E}\kern-.125emX}}
\begin{document}

\title{Multiple Speaker Separation from Noisy Sources in Reverberant Rooms using  Relative Transfer Matrix\\
\thanks{Thanks to Australian Research Council (ARC) Discovery Project Grant DP200100693 for funding.}
}

\author{
\IEEEauthorblockN{
Wageesha~N.~Manamperi$^\star \dagger$, Thushara~D.~Abhayapala$^\star$
}
\IEEEauthorblockA{\cr 
$^\star$ The Australian National University, Canberra, Australia \\
$^\dagger$ University of Moratuwa, Colombo, Sri Lanka 
}
}

\maketitle

\begin{abstract}
Separation of simultaneously active multiple speakers is a difficult task in environments with strong reverberation and many background
noise sources.  This paper uses the relative transfer matrix (ReTM), a generalization of the relative transfer function of a room, to propose a simple yet novel approach for separating concurrent speakers using noisy multichannel microphone recordings. 
The proposed method (i) allows multiple speech and background noise sources, (ii) includes reverberation, (iii) does not need the knowledge of the locations of speech and noise sources nor microphone locations and their relative geometry, and (iv) uses relatively small segment of recordings for training. We illustrate the speech source separation capability with improved intelligibility using a simulation study consisting of four speakers in the presence of three noise sources in a reverberant room. We also show the applicability of the method in a practical experiment in a real room.
\end{abstract}

\begin{IEEEkeywords}
Low SNR, multiple microphones, multiple sound sources, relative transfer matrix, speaker separation
\end{IEEEkeywords}


\section{Introduction}
\label{sec:1} 

Separation of multiple simultaneously active speakers in a room is a challenging acoustic signal processing problem. The aim is to recover the original speech signals or their equivalent version from the multichannel audio mixture which is contaminated by reverberation and 
background
noise.
Speaker separation will provide spatially selective listening for applications in augmented, virtual, and mixed reality.

Various approaches to speaker separation have been proposed using either single-channel or multichannel audio mixture, such as independent component
analysis (ICA) \cite{mitianoudis2003audio, sawada2004robust}, non-negative matrix factorization
(NMF) \cite{ozerov2009multichannel,acichocki2009nonnegativematrixandtensor}, beamforming \cite{gannot2017consolidated,markovich2009multichannel}, relative transfer function (ReTF) \cite{bates2020use}, and their combinations \cite{laufer2018source}, as well as machine learning techniques \cite{wang2018supervised,defossez2019music,nachmani2020voice,wang2018combining,gu2019neural,tesch2023multi,kalkhorani2024crossnet,subakan2021attention}. 
Nevertheless, most of the existing multi-talker separation methods, including the aforementioned ones, assumed the sparsity of the signals, or learned the discriminative patterns of speech, speakers, and background noise directly from rigorous training 
for a relative array geometry with or without given location information to extract the target speaker.

In \cite{abhayapala2023generalizing}, we introduced the concept of Relative Transfer Matrix (ReTM) which relates the signal received between two sets of microphone groups with respect to all active sources present in a room. Similar to the ReTF \cite{talmon2009relative}, the ReTM is independent of source signals but dependent on the spatial location of the sources and the environment. Recently, the ReTM of the noise sources has been used to propose a multi-channel speech denoising algorithm  \cite{kumar2024speech,wnmanamperiReTMdDrone}.
In this paper, we extend \cite{kumar2024speech},  for multi-speaker separation in a reverberant room under low signal-to-noise ratio (SNR) conditions. 
The key aspects of our method are as follows: (i) microphones are spatially distributed over the room, where microphone locations are unknown, (ii) microphones are divided into two groups, and (iii) neither source counting nor knowledge of the number of sources is required prior to the separation task. The method relies on estimating the ReTM with respect to all undesired sources (including noise) either semi-blindly or pre-training.  We also assume that the spatial properties of the environment, room, noise sources and speakers do not change with time. 
We show the performance of the proposed method using both simulation and real recordings in noisy environments. Also, we discuss merits of the proposed method with respect to recent work in speech separation.

\section{Problem Formulation and Relative Transfer Matrix}\label{sec:2}


In this section, we first formulate the problem of multichannel reverberant speech separation, and then review the relative transfer matrix (ReTM) method presented in \cite{abhayapala2023generalizing}, which divides the microphones into two groups.

\subsection{System Model}

Consider a reverberant environment with concurrently active $\mathcal{L}_S$ speech  and $\mathcal{L}_N$ background noise sources. Let $\mathcal{L}=\mathcal{L}_S+\mathcal{L}_N$. In the short time Fourier transform (STFT) domain, we denote $S_\ell^{(S)}(f,t) $, $\ell = {1,\cdots,\mathcal{L}_S}$ and $S_\ell^{(N)}(f,t)$, $\ell = {1,\cdots,\mathcal{L}_N}$ as the speech $(S)$ and background noise $(N)$ signals, respectively.

Let there be $Q$ arbitrary distributed microphones in the room. We divide them to two groups of microphones, $\{A\}$ and $\{B\}$ with $Q_A$ and $Q_B$ microphones, respectively ($Q = Q_A + Q_B$).
We denote $\mathbf{M}_A(f,t) $ and $\mathbf{M}_B(f,t) $  as the vector of received signals at microphone groups A and B, respectively.
Then the received signals at each microphone group in matrix form as
%
\begin{equation}\label{eqn:M_A}
    \mathbf{M}_A(f,t) = \mathbf{H}_A(f)\mathbf{S}(f,t) + \mathbf{V}_A(f,t),
\end{equation}
%
\begin{equation}\label{eqn:M_B}
    \mathbf{M}_B(f,t) = \mathbf{H}_B(f)\mathbf{S}(f,t) + \mathbf{V}_B(f,t)
\end{equation}
where 
$\mathbf{S}(f,t) = [S_1^{(S)}, \ldots, S_{\mathcal{L}_S}^{(S)}, S_1^{(N)}, \ldots, S_{\mathcal{L}_N}^{(N)}]^T$, and $\{\cdot\}^T$ is  the matrix transpose.
Here, $\mathbf{H}_A(f) \in \mathbb{C}^{Q_A \times \mathcal{L}}$ and $\mathbf{H}_B(f) \in \mathbb{C}^{Q_B \times \mathcal{L}}$ are the matrices with elements defined by the acoustic transfer functions.
The microphone thermal noise vector $\mathbf{V}_A(f,t) \in \mathbb{C}^{Q_A \times 1}$ and $\mathbf{V}_B(f,t) \in \mathbb{C}^{Q_B \times 1}$ are similarly defined.

The aim of this paper is to separate each individual speech signal $S_\ell^{(S)}(f,t)$, $\ell = {1,\cdots,\mathcal{L}_S}$, from the concurrent speakers $\mathcal{L}_S$ and the background noise sources $\mathcal{L}_N$.

\subsection{Background on the ReTM}

The ReTM, $\bm{\mathcal{R}}_{AB}(f)$, is define as in \cite{abhayapala2023generalizing} \begin{equation}\label{eqn:retm_hahb}
    \bm{\mathcal{R}}_{AB}(f) \triangleq \mathbf{H}_A(f) \mathbf{H}_B(f)^\dagger,
\end{equation}
where $(\cdot)^\dagger$ is Moore-Penrose inverse, assuming the validity, i.e., $Q_B \geq \mathcal{L}$. Thus, we can relate the received signal at group $\{A\}$ and $\{B\}$ using
%
\begin{equation} \nonumber 
     \mathbf{M}_A(f,t) - \mathbf{V}_A(f,t) = \bm{\mathcal{R}}_{AB}(f) \Bigl (\mathbf{M}_B(f,t) - \mathbf{V}_B(f,t) \Bigr).
\end{equation}
%
Note that $\bm{\mathcal{R}}_{AB}(f)$ 
is defined by the spatial properties of the sound sources such that it is independent of the sound source signals. In applications, the ReTM is invariant for a stationary environment.

The next section shows how to separate the desired speakers by estimating the ReTM of the undesired sources in practice.


\section{Multichannel Speaker Separation using ReTM  }\label{sec:3}

In the following, we extract the target speech of the $\ell^{th}$ speaker. Note that all speakers, $\ell = 1,\cdots,\mathcal{L}_S$, in the mixture can be similarly extracted. 

Let the target $\ell^{th}$ speech denotes as $S^{(S)}_\ell$ out of $L_S$ concurrent speakers. The rest of the undesired source signals including background noise source signals can be grouped as 
$\mathbf{\bar{S}}_\ell(f,t) = [S_1^{(S)} \ldots S_{\ell-1}^{(S)},S_{\ell+1}^{(S)}\ldots  S_{\mathcal{L}_S}^{(S)}, S_1^{(N)} \ldots S_{\mathcal{L}_N}^{(N)}]^T.$ We express the source signal vector
$ \mathbf{S}(f,t) = [S^{(S)}_\ell \, \mathbf{\bar{S}}_\ell^T]^T$.

%
Let $\bm{\mathcal{R}}_{AB}^{({\ell})}(f)$ be the ReTM of the combination of all sound sources except the $\ell^{th}$ target source. 
Here, we assume a specific scenario of two time segments with known segment boundaries as (i) undesired sources-only segment ($\mathcal{T}_1$) (to be used to estimate ReTM), and (ii) a sound segment ($\mathcal{T}_2$) with a mixture of active sources including the desired source. We use the covariance matrices-based approach as in \cite{abhayapala2023generalizing} to blindly estimate the $\bm{\mathcal{R}}_{AB}^{({\ell})}(f)$ using  $\mathcal{T}_1$ segment of the microphone recording in which the target $\ell^\text{th}$ speaker is inactive. Hence, 
\begin{equation}\label{eqn:retm_rho}
\setlength{\abovedisplayskip}{-0.7pt}
\setlength{\belowdisplayskip}{-0.7pt}
     \bm{\mathcal{R}}_{AB}^{({\ell})}(f) \approx \bm{\mathcal{P}}_{AA}^{({\ell})}(f) \,  \Big( \bm{\mathcal{P}}_{BA}^{({\ell})}(f) \Big)^\dagger,
\end{equation}
with $\bm{\mathcal{P}}_{AA}^{({\ell})}(f) \triangleq E\{\mathbf{M}_A^{({\ell})}(f,t) \mathbf{M}_A^{({\ell})^*}(f,t)\}$, and $\bm{\mathcal{P}}_{BA}^{({\ell})}(f)  \triangleq E\{\mathbf{M}_B^{({\ell})}(f,t) \mathbf{M}_A^{({\ell})^*}(f,t)\}$, where $E\{\cdot\}$ denotes the expectation which can be obtained by averaging across the time frames. Although blind estimation of ReTM is used in this paper, the proposed method can also be implemented by pre-trained or semi-blind estimation (e.g., conference rooms with pre-arranged settings).

We remove all undesired speech sources and background noise from the group $\{A\}$ microphone signals $\mathbf{M}_A(f,t)$ by multiplying group $\{B\}$ microphone signals $\mathbf{M}_B(f,t)$ by $\bm{\mathcal{R}}_{AB}^{({\ell})}(f)$ and subtracting as 
\begin{equation} \label{eqn:desiredspeechl}
    \hat{\mathbf{S}}_\ell(f,t) \triangleq  \mathbf{M}_A(f,t) -  \bm{\mathcal{R}}_{AB}^{({\ell})}(f) \mathbf{M}_B(f,t),
\end{equation}
where $\mathbf{\hat{S}}_\ell(f,t)$ is a $ Q_A \times 1$ vector consists $Q_A$ copies of estimated target speech signal $S^{(S)}_\ell$. 
%
For convenience, we omit the dependency of time ($t$) and frequency ($f$) in the rest of this section. 
Using \eqref{eqn:M_A} and \eqref{eqn:M_B} in \eqref{eqn:desiredspeechl}, we obtain
%
\begin{equation} 
  \hat{\mathbf{S}}_\ell = [\mathbf{H}_A  - \bm{\mathcal{R}}_{AB}^{({\ell})} \mathbf{H}_B ] [S^{(S)}_\ell \, \mathbf{\bar{S}}_\ell^T]^T + \mathbf{V}_A - \bm{\mathcal{R}}_{AB}^{({\ell})}  \mathbf{V}_B.
\end{equation}
%
Let $\mathbf{h}^{(\ell)}_A$ and $\mathbf{h}^{(\ell)}_B$ be the acoustic transfer function vectors from the $\ell^{\text{th}}$ speech source to group $\{A\}$ and $\{B\}$ microphones, respectively. Also, let $\mathbf{\bar{H}}^{(\ell)}_A$ and $\mathbf{\bar{H}}^{(\ell)}_B$ be the acoustic transfer function matrices from all other sources except the $\ell^{\text{th}}$ speech source to group A and B microphones, respectively. Thus, 

\setlength{\abovedisplayskip}{-0pt}
\setlength{\belowdisplayskip}{-2pt}
\begin{align}
\hat{\mathbf{S}}_\ell 
 &= [\mathbf{h}^{(\ell)}_A  - \bm{\mathcal{R}}_{AB}^{({\ell})} \mathbf{h}^{(\ell)}_B] S^{(S)}_\ell \nonumber \\
 &+ [\mathbf{\bar{H}}^{(\ell)}_A  - 
 {\bm{\mathcal{R}}_{AB}^{({\ell})} \mathbf{\bar{H}}^{(\ell)}_B}
 ]
 \mathbf{\bar{S}}_\ell 
 \, + \,  \mathbf{V}_A - \bm{\mathcal{R}}_{AB}^{({\ell})}  \mathbf{V}_B  \nonumber \\
 &= [\underbrace{\mathbf{h}^{(\ell)}_A  - \bm{\mathcal{R}}_{AB}^{({\ell})} \mathbf{h}^{(\ell)}_B}_{\text{distortion}}] S^{(S)}_\ell + \underbrace{\mathbf{V}_A - \bm{\mathcal{R}}_{AB}^{({\ell})}  \mathbf{V}_B}_{\text{residual thermal noise}}. \label{eqn:retm_sss}
\end{align}


Equation \eqref{eqn:retm_sss}, along with an accurate estimate of $\bm{\mathcal{R}}_{AB}^{({\ell})}$, $\ell=1,\ldots,\mathcal{L}_S$, provides a complete separation of all speakers from the mixture.
However, \eqref{eqn:retm_sss} is a `distorted' version of the target speech signal $S^{(S)}_{\ell}$ in terms of the room transfer matrix of both speech and background noise sources with some microphone thermal noise. We find that the impact of the thermal noise term in \eqref{eqn:retm_sss} is minimal in our experiments. 





\section{Experimental Evaluation}
\label{sec:4}

This section shows the implementation of the proposed separation method under low (below $0$ dB) signal-to-background noise ratio (SNR) conditions in a reverberant room using both simulation and real recordings.

\subsection{Simulated Environments}\label{sec:4.1}
%
We utilize an open-source toolbox \cite{rirgen} to model the room impulse response (RIR) from the sound sources to irregularly distributed microphones in a  $6 \times 7 \times 3$ m rectangular room ($T_{60} = 500$ ms). We consider four speech sources, three background noise sources, and $27$ microphones. Four speaker locations are: (1): (3 m, 4.5 m, 1.2 m), (2): (3 m, 2.5 m, 1.2 m), (3): (4.5 m, 3.5 m, 1.2 m), and (4): (1.5 m, 3.5 m, 1.2 m) and three background noise sources locations are: (1): (2 m, 0.9 m, 1.8 m), (2): (1 m, 6 m, 2.5 m), and (3): (3.5 m, 5 m, 1.8 m) with respect to the origin position in the left corner of the room. We convolve the speech sources RIRs with both male and female speech utterances from the TIMIT dataset \cite{garofolo1993timit} and noise sources RIRs with wall air-conditioner noise, vacuum noise, and music signal. The received signals are down-sampled to $16$ kHz and ranged from $0$ to $-20$ dB SNR of background noise and added with $60$ dB SNR of white Gaussian noise at each microphone. Here, we define the SNR by averaging SNR at each receiver over all $27$ receivers.

We assume this undesired sound sources-only signal $\mathbf{\bar{S}}_\ell$ has been obtained when each $\ell^{th}$ target speaker $\ell \in \{1,\cdots,\mathcal{L}_S\}$ is inactive and use a $60$ second segment of this recording for $\bm{\mathcal{R}}_{AB}^{({\ell})}$ training. Then, the recordings are short-time-Fourier-transformed with an 8192-point window size that was long relative to the length of the RIR to satisfy the multiplicative transfer function \cite{avargel2007multiplicative}. We assign $Q_A \in \{5,7,10\}$ varying number of receivers to group $\{A\}$, and $Q_B = 17$ a fixed number of receivers to group $\{B\}$.

We use two qualitative measures to assess the separation performance: (i) Signal-to-Interference Ratio (SIR), and (ii) Signal-to-Distortion Ratio (SDR), using the BSS-Eval toolbox \cite{vincent2006performance}. We also measure speech intelligibility through Short-Time Objective Intelligibility (STOI) \cite{taal2011algorithm}. The SNR is defined with respect to all sources in the mixture, whereas SIR is defined considering one speaker as the target signal and the rest of the interfering speakers as the noise signal.

We first evaluate performance in a varying number of $Q_A$ microphones with fixed $Q_B$ at $-15$ dB SNR level in Table~\ref{table:sss_qavarying}.
We observe that both group A sizes of $7$ and $10$ yield the best results, however, $Q_A = 7$ obtains the highest SDR. The total number of sources in the mixture is $\mathcal{L} = 7$, suggesting a lower bound for the groups $\{A,B\}$ size for the improved separation capabilities. The results confirm that with the higher $Q_A, Q_B ( \geq \mathcal{L})$ accurately separated the speech signals in noisy reverberant environments.

\begin{table}[!ht]
\caption{Separation result of the proposed method at microphone channel $1$ of $Q_A$ with a varying number of group $\{A\}$ receivers and fixed group $\{B\}$ receivers at $-15$ dB SNR. O/P denotes the output or the separated speech signal.} 
\label{table:sss_qavarying}
\vskip3pt
\centering
\begin{tabular}{|c|c|c|c|c|}
\hline
Speaker  &   & STOI  &  SIR  & SDR  \\
 Num.  &   & ($\%$)  & (dB) & (dB) \\
\hline
\multirow{2}{*}{1} &  Unprocessed &   37.35 & 1.40 &  - \\
 &  O/P$_{\{Q_A=5\}}$ &  84.79 & 15.41 & 7.22 \\
 &  O/P$_{\{Q_A=7\}}$ &  92.16 & 25.37 & 9.49 \\
 &  O/P$_{\{Q_A=10\}}$ &  91.83 & 31.23 & 7.53 \\
\hline
\multirow{2}{*}{2} &  Unprocessed &   23.85 & -6.82 &  - \\
 &  O/P$_{\{Q_A=5\}}$ &  59.78 & 9.94 & 0.86 \\
 &  O/P$_{\{Q_A=7\}}$ &  73.02 & 19.09 & 3.47 \\
 &  O/P$_{\{Q_A=10\}}$ &  72.61 & 26.68 & 2.98 \\
\hline
\multirow{2}{*}{3} &  Unprocessed &   20.79 & -14.35 &  - \\
 &  O/P$_{\{Q_A=5\}}$ &  40.65 & 0.82 & -7.21 \\
 &  O/P$_{\{Q_A=7\}}$ &  66.67 & 10.96 & -1.86 \\
 &  O/P$_{\{Q_A=10\}}$ &  69.76 & 21.83 & -2.83 \\
\hline
\multirow{2}{*}{4} &  Unprocessed &   29.59 & -5.69 &  - \\
 &  O/P$_{\{Q_A=5\}}$ &  60.77 & 9.56 & -0.15 \\
 &  O/P$_{\{Q_A=7\}}$ &  71.50 & 19.65 & 1.45 \\
 &  O/P$_{\{Q_A=10\}}$ & 71.37 & 23.89 &  -0.16 \\
\hline
\end{tabular}
\end{table}

\begin{figure}[h!]
\centering
\includegraphics[scale=0.4]{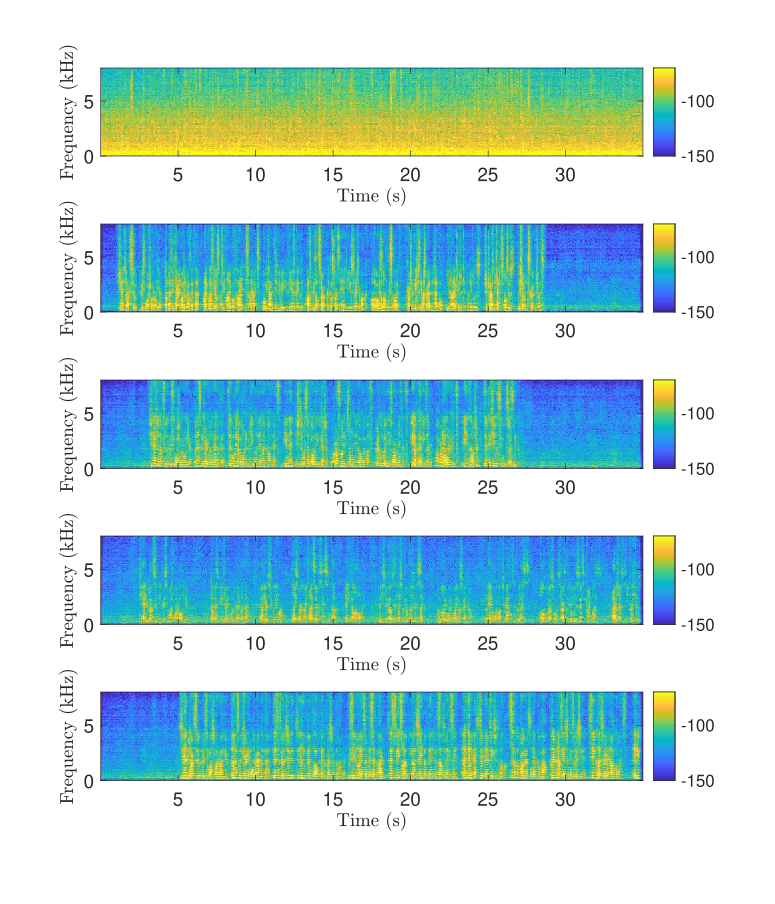}
\vspace*{-1.2cm}
\caption{ Spectrogram plots of microphone recordings and separated speech at microphone channel $1$ of $Q_A$ at $-15$ dB SNR level.}\label{fig:spectro_snr}
\end{figure}

\addtolength{\textfloatsep}{-0.2in}

We will now examine the spectrogram plots of the mixture signal, separated speech signal outputs at the microphone $1$ of $Q_A = 10$ in Fig.~\ref{fig:spectro_snr}. We observe that the input signal is contaminated by background noise, however, the separated output for each speaker is less noisy. We share a link to the mixture signal and the separation speech signals to listen.\footnote{https://github.com/wnilmini/SS-ReTM}

\begin{figure*}[t]
    \centering
    \subfigure[]{\includegraphics[width=0.32\textwidth]{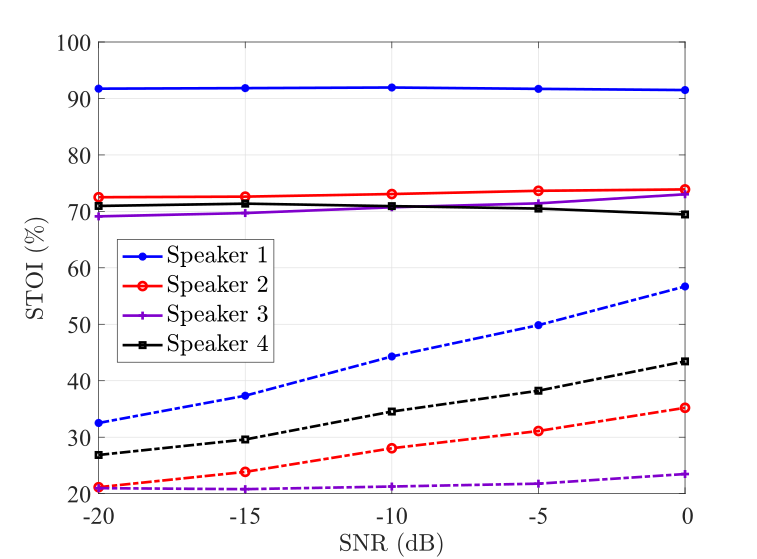}} 
    \subfigure[]{\includegraphics[width=0.32\textwidth]{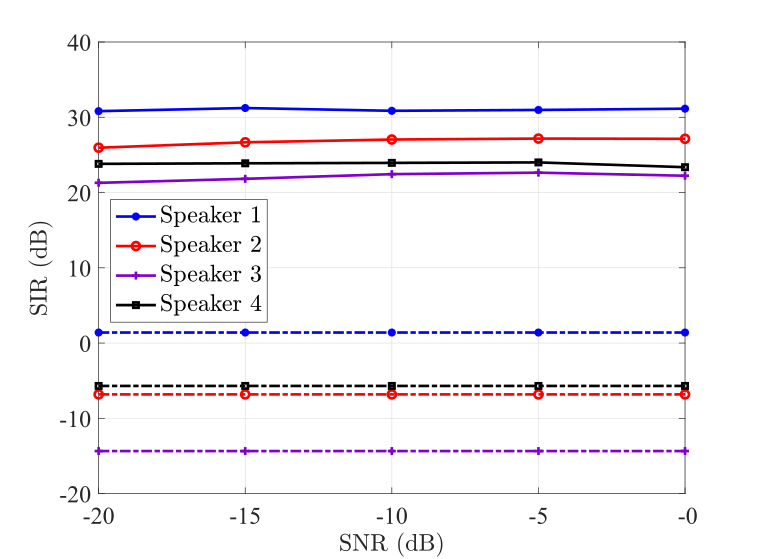}}
    \subfigure[]{\includegraphics[width=0.32\textwidth]{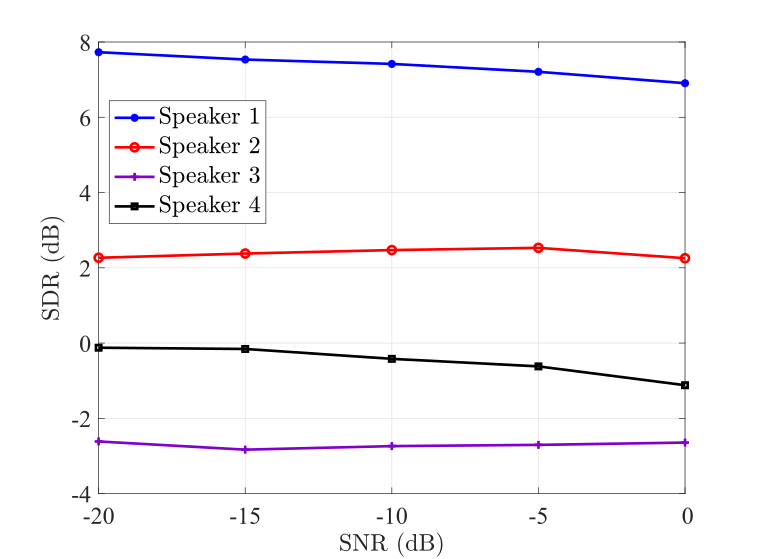}}
    \vspace*{-0.2cm}
    \caption{(a) STOI, (b) SIR, and (c) SDR performance of the proposed method as a function of the SNR over four speakers. The dotted line shows the input, whereas the solid line shows the output measures of the corresponding target speaker.}
    \label{fig:sss_eval}
\end{figure*}

Next, we examine the performance of the proposed algorithm with different SNR levels in Fig.~\ref{fig:sss_eval}. Given that the proposed method works better with  more microphones in groups A and  B than the number of sound sources, we set $Q_A = 10$ and $Q_B = 17$ with varying SNR levels from $0$ dB to $-20$ dB.  

Results of Fig.~\ref{fig:sss_eval} show that the proposed method is able to successfully separate all speakers in the mixture, especially with the different input speaker configurations, i.e., STOI, SIR, at low SNR levels. Three key observations are made from Fig.~\ref{fig:sss_eval}. First is in terms of intelligibility improvement, where we observe the highest output STOI values for the same speaker with the highest input STOI while the rest achieve comparable results. The second observation is on separation results, where we observe improved SIR, and SDR for individual speakers with a performance gap that the first speaker is dominated which has the highest SIR value if four speakers are with different SIR levels. Our third observation is that the speaker separation results are seen to be nearly constant for each speaker over the low input SNR conditions as focusing constant input SIR of the individual speaker.

\begin{figure}[h!]
\centering
\includegraphics[scale=0.35]{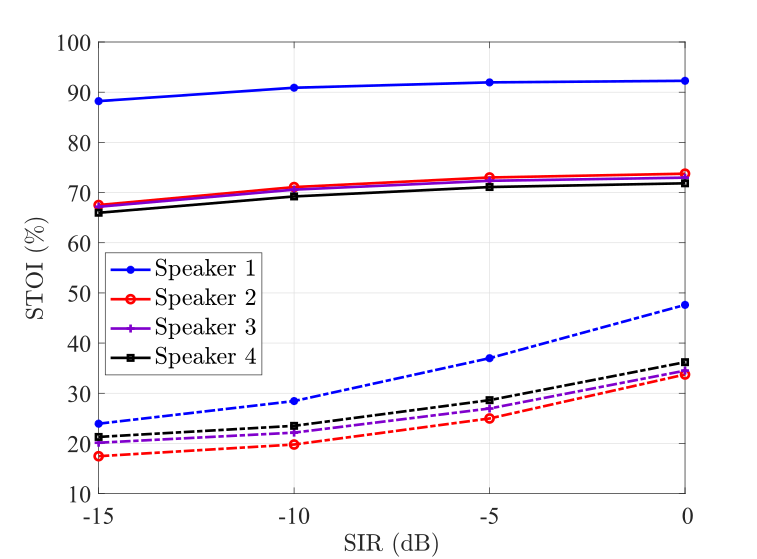}
\vspace*{-0.25cm}
\caption{ STOI performance as a function of the input SIR over four speakers at $-15$ dB SNR level.
}\label{fig:stoi_sir}
\end{figure}

Finally, we consider intelligibility performance with different SIR levels from $-15$ dB to $0$ dB and a fixed SNR level of $-15$ dB to explain both the second and the third observations on speaker separation. In Fig.~\ref{fig:stoi_sir} we present four scenarios with the assumption that the target speaker's SIR levels are varied given that the other three speakers and the background noise sources in the mixture are kept the same. This is similar to Fig.~\ref{fig:sss_eval}(a), where the first speaker with the highest SIR among all four speakers in the mixture is well-separated by the proposed method achieving the highest STOI performance. It can be seen in Fig.~\ref{fig:stoi_sir} that the proposed method has indistinguishable intelligibility performance for the other three speakers in the mixture. Note that the algorithm's robustness to room reverberation is not included in Section~\ref{sec:4}, since it provided similar performance.

\subsection{Real-life Environments}

\begin{figure}
    \centering
    \includegraphics[angle=180,width=0.8\linewidth]{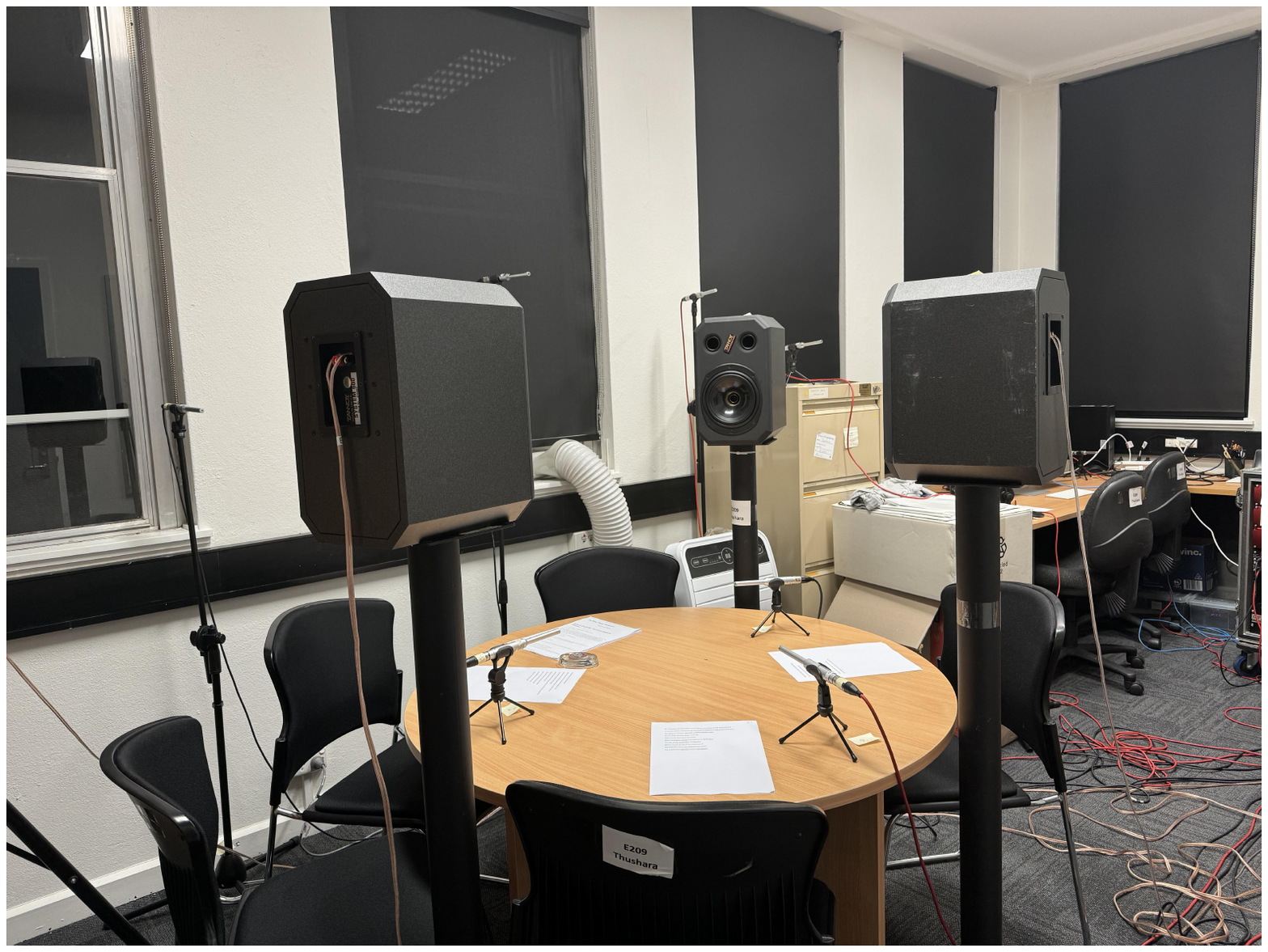}
    \caption{Experimental setup for real measurements in a reverberant room, where two speakers are sat around a table and 15 microphones are distributed across the room.}
    \label{fig:real_rec}
\end{figure}

Now, we conduct a preliminary study of this work in a real-life scenario with highly overlapped speakers. The real recordings are measured in an office room at the Australian National University with dimensions  $2.94 \times 4.4 \times 3.04$ m, and $T_{60} \approx 410$ ms. We consider $2$ speakers, $2$ background noise sources (fan and room air cooler), and $15$ randomly distributed microphones over the room (as shown in Fig.~\ref{fig:real_rec}). We assigned $5$ microphones (channels $\{3,4,10,12,15\}$) to group A, and $10$ microphones (channels $\{1,2,5:9,11,13,14\}$) to group B. We examined the proposed method's performance in separating the desired speakers inside a real room. We share a link to the audio files.\footnote{https://zenodo.org/records/15009085} By listening to the output signals, we observe that the leakage of the interfering sources into the separated desired speakers. However, the desired speakers are clearly audible to understand their utterances. Note that the separation results are inferior to the results obtained in simulations. We attribute this discrepancy can be modified by using Wiener filtering techniques to enhance the separation performance, which we will continue to discuss in future work.

\subsection{Comparison Methods}


The machine learning-based methods \cite{ozerov2009multichannel,acichocki2009nonnegativematrixandtensor,nachmani2020voice,wang2018combining,gu2019neural,tesch2023multi,kalkhorani2024crossnet,subakan2021attention} for audio speaker separation consider data-driven approaches and exploit either speech features or sparsity in their models/algorithms, perhaps both in a supervised \cite{nachmani2020voice,tesch2023multi} or unsupervised \cite{ozerov2009multichannel} manner whereas in the proposed algorithm only requires speaker spatial location or environment-based blind (or semi-blind) training  among the microphone groups. Both DNN-based methods and NMF-based methods require greater computational power which aggravates with the number of microphone channels in the mixture whereas our proposed method alleviates the computational cost of multichannel speaker separation using one to two minutes of the ReTM training. Moreover, DNN-based methods rely on numerous assumptions about the acoustic environment or the background noise sources and essential rigorous training for specific microphone array structures and/or a given number of speakers in the mixture. However, the proposed method is independent of the microphone placement in the specific acoustic scenario, however, requires a more spatially diverse microphone configuration to learn the ReTM. Furthermore, we note that our algorithm is independent of the source signals, and the number of sources prior to the separation task as well as, speech-based features like pitch and sparsity which are unreliable under noisy environments.

As our method requires multiple microphones (groups A and B should have microphones more than the number of sound sources in the mixture), it is not direct and easy to compare fairly with existing datasets and tasks to reproduce with baseline methods with appropriate configurations. But in the future, we seek to have a fair comparison with the baseline speaker separation methods. 

For now, we will present the performance evaluation of the DNN-based method, SepFormer \cite{subakan2021attention}. To fairly evaluate the algorithms, we use the same settings given in \ref{sec:4.1} with SNR $= 0$ dB.  Table \ref{table:basline} depicts the speaker separation results for $\mathcal{L}_S = 3$. The results confirm that both methods effectively separate the speaker with the highest loudness. Although 'SepFormer' exhibits slightly higher performance than the input signal, it fails to effectively separate the concurrent speakers when their loudness levels vary, whereas our method successfully achieves this separation. Note that the results for low SNR levels are not reported here, since it is severely impacted by the noise. We share a link to the audio files.\footnote{https://github.com/wnilmini/SS-ReTM/tree/main/Baseline}

\begin{table}[!ht]
\caption{Performance comparison with the baseline.} 
\label{table:basline}
\vskip3pt
\centering
\begin{tabular}{c|c|c|c|c}
\hline
Spk. No. &  & STOI ($\%$)  &  SIR (dB) & SDR (dB)  \\
\hline
\hline
\multirow{2}{*}{1} &  Unprocessed &   63.05 & 3.92 &  - \\
 & SepFormer \cite{subakan2021attention} &  74.30 & 7.20 & 4.68 \\
 &  Our Method &  $\bm{92.59}$ & $\bm{34.87}$ & $\bm{8.11} $\\
\hline
\multirow{2}{*}{2} &  Unprocessed &   39.63 & -5.71 &  - \\
 & SepFormer \cite{subakan2021attention} &  38.63 & 2.89 & -5.96 \\
 & Our Method &  $\bm{75.61}$ & $\bm{29.05}$ & $\bm{3.03}$\\
\hline
\multirow{2}{*}{3} &  Unprocessed &   32.58 & -10.77 &  - \\
 &  SepFormer \cite{subakan2021attention} &  34.10 & -4.62 & -14.20 \\
 & Our Method &  $\bm{72.60}$ & $\bm{26.17}$ & $\bm{-1.15}$ \\
\hline
\end{tabular}
\end{table}


\section{ Conclusion}\label{sec:6}

In this paper, we have presented a method for multichannel reverberant speech separation in low SNR levels. The method exploits a unique spatial signature of the sound sources distributed throughout the environment using the ReTM. The ReTMs of the undesired sound sources are blindly estimated when the desired speaker is not active to separate the individual speakers. The simulation results have shown that the proposed algorithm provides good speaker separation performance with an increased number of microphone channels in both groups $\{A, B\}$ than the total number of sound sources in the mixture. In the future, we will conduct a fair comparison of the proposed speaker separation method with the state-of-the-art methods and further examine the performance of the multiple-speaker separation in real-life scenarios.


\bibliographystyle{IEEEtran}
\bibliography{paper}

\end{document}